\documentstyle[preprint,aps,eqsecnum]{revtex}
\begin{document}
\draft
\preprint{Alberta Thy-37-94, hep-th/9412041}
\title{Finite particle creation in 3+1 de Sitter space}
\author{D.\ J. Lamb\cite{email}, A.\ Z. Capri and
M. Kobayashi\cite{address} }
\address{ Theoretical Physics Institute, \\
Department of Physics,
University of Alberta,\\
Edmonton, Alberta T6G 2J1, Canada}
\date{\today}
\maketitle
\begin{abstract}
 In this paper we calculate the particle creation as seen by a stationary
observer in 3+1 de Sitter space. This particle creation is calculated using an
observer dependent geometrically based definition of time which is used to
quantize a field on two different spacelike surfaces. The Bogolubov
transformation relating these two quantizations is then calculated and the
resulting particle creation is shown to be finite.
\end{abstract}

\pacs{03.70}

\section{Introduction}
In this paper we calculate the particle creation as seen by a stationary
observer in 3+1 de Sitter space. This particle creation is calculated by
looking
at the Bogolubov transformation relating the observer's different definitions
of particle states on two different spacelike hypersurfaces. The definition of
particle states used is
that proposed by Capri and Roy \cite{Caproy92} and is equivalent to the
definition proposed by Massacand and Schmid \cite{eth}. This definition of
particle states uses a coordinate independent definition of time which one uses
to decompose the field into
positive and negative frequency parts. This time is defined as being normal to
the spacelike geodesic hypersurface which intersects the observer's worldline
orthogonally. In this way the spacetime is spanned by geodesics. If there is a
geodesically complete coordinatization for the spacetime this is the
coordinatization that will be picked out by this definition of time. In de
Sitter
space this implies that the radial coordinate is compact even though the
cooridinatization we start with would not suggest this. It is this compact
coordinatization that allows us to eventually integrate by parts the expression
for the total particle production and show that it is finite. Similar results
were obtained in an earlier paper \cite{us} for a $1+1$ dimensional model which
was compact in space.

The particle production is shown to be finite as the Bogolubov $\beta(N,N',l)$
coefficient drops off faster than any inverse power of $N$ or $N'$. If this
drop off is actually an exponential then the particle production would be
consistent with a thermal distribution which is what is expected for the large
momenta limit. This finite particle creation agrees with the analysis presented
in Fulling's book for expanding isotropic universes \cite{full}.

This calculation is not a calculation of the Bogolubov transformation relating
two different coordinatizations of, the same spacetime or, different portions
of the same spacetime.

\section{The Model}
   We start with the following coordinatization of de Sitter space,
\begin{equation}
ds^2=dT^2-e^{\lambda T}\left((dX^1)^2+(dX^2)^2+(dX^3)^2 \right)
\label{2.1}
\end{equation}
To calculate the coordinates which provide the foliation mentioned in the
introduction we must first calculate the geodesic equations. The first
integrals of the geodesic equations are,
\begin{equation}
\frac{dX^i}{ds}=c^ie^{-\lambda T}\ \ \ {\rm and} \ \ \
\frac{dT}{ds}=\sqrt{\epsilon + e^{-\lambda T} {\bf c}^2}
\label{2.2}
\end{equation}
where  $i=1$ to $3$ and $\epsilon = \pm 1$ depending on whether the geodesic is
timelike or spacelike respectively. The preferred coordinates on the
hypersurface of instantaneity are constructed using a $4$-bein of orthonormal
basis vectors based at $P_0$, the observer's position. These vectors are
chosen to be,
\begin{eqnarray}
e^0_{\mu}(P_0)&=&(1,0,0,0)\ \ \ e^1_{\mu}(P_0)=(0,e^{-\frac{\lambda
T_0}{2}},0,0) \nonumber \\
e^2_{\mu}(P_0)&=&(0,0,e^{-\frac{\lambda T_0}{2}},0)
\ \ \ e^3_{\mu}(P_0)=(0,0,0,e^{-\frac{\lambda T_0}{2}})
\label{2.3}
\end{eqnarray}
In this way $e^0_{\mu}(P_0)$ is tangent to the worldline of an observer which
is
stationary with respect to the coordinates of (\ref{2.1}).
To construct a spacelike geodesic which is orthogonal to the observer's
worldline it is required that
\begin{equation}
\frac{dT}{ds}|_{P_0}=0\ \ \ {\rm which\ \ \ \ implies} \ \ \ {\bf
c}^2=e^{\lambda T_0}.
\label{2.4}
\end{equation}
The preferred coordinates on the spacelike hypersurface are chosen to be
Riemann coordinates based on the observer's position
$P_0=(T_0,X^1_0,X^2_0,X^3_0)$. The coordinates are constructed using the point
$P_1=(T_1,X^1_1,X^2_1,X^3_1)$ which is the point at which a timelike geodesic
``dropped" from an arbitrary point
$P=(T,X^1,X^2,X^3)$ intersects the spacelike hypersurface orthogonally. The
Riemann coordinates $\eta^\alpha$ of the point $P_1$ are given by,
\begin{equation}
s_s p^{\mu}=\eta^{\alpha}e^{\mu}_{\alpha}(P_0)
\label{2.5}
\end{equation}
where $s_s$ is the distance along the geodesic $P_0-P_1$ and $p^\mu$ is the 
vector tangent to the geodesic connecting $P_0$ to $P_1$, at $P_0$.
These equations can be solved for the coordinates $\eta^\alpha$ using the
orthogonality
of $p^\mu$ to $e_0(P_0)$ and the identity $e_\alpha^\mu e_{\beta \mu}=
\eta_{\alpha \beta}$ (Minkowski metric) to give,
\begin{equation}
\eta^0=s_s p^\mu e_\mu^0(P_0) \ \ \ \ \eta^i=-s_s p^\mu e_\mu^i(P_0)
\label{2.55}
\end{equation}
The surface of instantaneity is then just the surface $\eta^0=0$ and the
preferred spatial coordinates are given by,
\begin{equation}
x^i = s_s c^i e^{\frac{\lambda T_0}{2}}
\label{2.6}
\end{equation}
The preferred time coordinate $t$ of an arbitrary point $P$ is then given by
the geodesic distance along the timelike geodesic connecting $P$ to $P_1$. This
timelike geodesic is also determined by (\ref{2.2}) with a different set of
constants $b^i$ and $\epsilon=1$. The condition that this timelike geodesic is
orthogonal to the
spacelike hypersurface is,
\begin{equation}
\sqrt{e^{\lambda(T_0-T_1)-1}}\sqrt{{\bf c}^2e^{-\lambda T_1}+1}={\bf
c}\cdot{\bf b}e^{-\lambda T_1}
\label{2.7}
\end{equation}
There is an arbitrary choice involved in how one solves these two equations for
the
constants ${\bf b}$ and ${\bf c}$. This freedom can be understood as the
ability to rotate the hypersurface of instantaneity through a reparametrization
of the surface.
The choice which we  make for reasons of calculational simplicity is that
\begin{equation}
b^i = \sqrt{1-e^{-\lambda(T_0-T_1)}} c^i.
\label{2.8}
\end{equation}
At this point it convenient to to introduce the variable $r$,
\begin{equation}
r^2= {\bf x}\cdot {\bf x} = s_s^2 \frac{{\bf c}\cdot {\bf c}}{e^{\lambda
T_0}}=s_s^2.
\label{2.9}
\end{equation}
 We can now calculate the metric in terms of the preferred coordinates $(t,{\bf
x})$ by calculating $(T(t,x^i),{\bf X}(t,{\bf x}))$.
\begin{equation}
X^i=X_0^i + \int_{T_0}^{T_1}dT\frac{c^i e^{-\lambda
T}}{\sqrt{e^{\lambda(T_0-T)}-1}} + \int_{T_1}^{T}dT\frac{b^i e^{-\lambda
T}}{\sqrt{1+{\bf b}^2e^{-\lambda T}}}
\label{2.10}
\end{equation}
We also need to calculate $t$ and $s_s$,
\begin{equation}
s_s= \int_{T_0}^{T_1}\frac{dT}{\sqrt{e^{\lambda(T_0-T)}-1}}
\label{2.11}
\end{equation}
\begin{equation}
t=\int_{T_1}^{T}\frac{dT}{\sqrt{1+{\bf b}^2e^{-\lambda T}}}
\label{2.12}
\end{equation}
One can now obtain the coordinate transformations,
\begin{eqnarray}
e^{\frac{\lambda}{2}(T-T_0)}=\cosh(\frac{\lambda t}{2})\cos(\frac{\lambda
r}{2}) +
\sinh(\frac{\lambda t}{2}) \nonumber \\
\frac{\lambda}{2}(X^i-X_0^i)e^{\frac{\lambda}{2}T}=
\frac{x^i}{r}\cosh(\frac{\lambda t}{2})\sin(\frac{\lambda r}{2}).
\label{2.13}
\end{eqnarray}
We can see here by looking at a particular $t=$constant surface that the
range of $r$ is now compact and the range $0\leq \frac{\lambda r}{2} < \pi $
covers the entire manifold which was covered by the original coordinates
$(T,{\bf X})$.
It is now easy to put the preferred coordinates into polar form,
\begin{eqnarray}
x^1&=&r \sin(\theta)\sin(\phi) \nonumber \\
x^2&=&r \sin(\theta)\cos(\phi) \nonumber \\
x^3&=&r \cos(\theta)
\label{2.14}
\end{eqnarray}
In terms of these preferred coordinates the metric is,
\begin{equation}
ds^2=dt^2-\cosh^2(\frac{\lambda t}{2})\left(
dr^2+\frac{4}{\lambda^2}\sin^2(\frac{\lambda r}{2}) \left(d\theta^2
+\sin^2(\theta)d\phi^2 \right)\right)
\label{2.15}
\end{equation}
This result is of course no surprise to anyone familiar with different
coordinatizations of de Sitter space, given that the space was being
coordinatized in terms of geodesics. The point here is not what the final form
of the metric is as much as how these
transformations will change as our observer moves to a different point and the
entire construction is repeated.

\section{Modes and Initial Conditions}
  In the coordinates constructed above, the minimally coupled massless Klein
Gordon equation is,
\begin{equation}
\partial^2_t \phi +  \frac{1}{\sqrt{g}}\partial_t\left( \sqrt{g}\right)
\partial_t \phi + \frac{1}{\sqrt{g}}\partial_i\left( \sqrt{g}g^{ij}\right)
\partial_j \phi = 0
\label{3.1}
\end{equation}
where $|g|$ and the $g^{ij}$ can be read off from (\ref{2.15}).
To quantize a scalar field on the $t=0$ surface we now define positive the
frequency modes as those which satisfy the initial conditions,
\begin{equation}
\phi^{+}_{N l n}=A_{N l n}(0,r,\theta,\phi)\ \ \ \  {\rm and }\ \ \
\partial_t\phi^{+}_{N l n}|_{t=0}= -i\omega_N(0) A_{N l n}(0,r,\theta,\phi).
\label{3.2}
\end{equation}
Where $A_{N l n}(0,r,\theta,\phi)$ are the instantaneous eigenmodes of the
spatial part of the Laplace-Beltrami operator, and $\omega_N(t)^2$ are the
corresponding eigenvalues,
\begin{equation}
\frac{1}{\sqrt{g}}\partial_i\left( \sqrt{g}g^{ij}\right)\partial_j
A_{N l n}(t,r,\theta,\phi) = \omega_N(t)^2 A_{N l n}(t,r,\theta,\phi).
\label{3.3}
\end{equation}
Henceforth we write $\omega_N$ for $\omega_N(0)$.
\begin{equation}
\omega_N\equiv \omega_N(0)=\sqrt{\frac{\lambda^2}{4}N(N+2)}
\label{3.35}
\end{equation}
The differential
equations (\ref{3.1}) and (\ref{3.3}) must now be solved and the appropriate
initial conditions imposed.
The positive frequency solution to these differential equations which satisfies
the correct initial conditions as just stated is,
\begin{eqnarray}
\phi^+_{N l n}(t,r,\theta,\phi) =
F_{N l}Y_{l n}(\theta,\phi)\sin^l(\frac{\lambda r}{2}){\rm
C}_{N-l}^{l+1}\left[\cos(\frac{\lambda r}{2}) \right] {\rm sech}^{\frac{3}{2}}
(\frac{\lambda t}{2})\times \nonumber \\ \left( L
P_{\frac{1}{2}+N}^{\frac{3}{2}}\left[\tanh(\frac{\lambda t}{2})\right] +  M
Q_{\frac{1}{2}+N}^{\frac{3}{2}}\left[\tanh(\frac{\lambda t}{2})\right]
\right)
\label{3.4}
\end{eqnarray}
where
\begin{eqnarray}
L&=&-{{(2+N)\,{ \lambda}\,{Q}_{-{1\over 2} + {N}}^{{3\over 2}}(0)
        - 2\,i\,{Q}_{{1\over 2} + {N}}^{{3\over 2}}(0)\,
        \omega_{N}}\over
     {{ \lambda}\,\left( 2 + {N} \right) \,
       \left( - {P}_{{1\over 2} + {N}}^{{3\over 2}}(0)\,
            {Q}_{-{1\over 2} + {N}}^{{3\over 2}}(0)  +
         {P}_{-{1\over 2} + {N}}^{{3\over 2}}(0)\,
          {Q}_{{1\over 2} + {N}}^{{3\over 2}}(0) \right) }}
\nonumber \\
M&=&{{(2+N)\,{ \lambda}\,{P}_{-{1\over 2} + {N}}^{{3\over 2}}(0)  -
2\,i\,{P}_{{1\over 2} + {N}}^{{3\over 2}}(0)\,
       \omega_{N}}\over
    {{ \lambda}\,\left( 2 + {N} \right) \,
      \left( - {P}_{{1\over 2} + {N}}^{{3\over 2}}(0)\,
           {Q}_{-{1\over 2} + {N}}^{{3\over 2}}(0)   +
        {P}_{-{1\over 2} + {N}}^{{3\over 2}}(0)\,
         {Q}_{{1\over 2} + {N}}^{{3\over 2}}(0) \right) }}
 \nonumber \\
F_{N l}&=& {{{2^{{1\over 2} + l}}\,{\sqrt{1 + { N}}}\,{ \Gamma}(1 + l)\,
      {\sqrt{{ \Gamma}(1 - l + { N})}}}\over
    {{\sqrt{\pi }}\,{\sqrt{{ \Gamma}(2 + l + { N})}}}},
\label{3.5}
\end{eqnarray}
$C_m^n[x]$ are Gegenbauer polynomials and $P^m_n[x]$ and $Q^m_n[x]$ are
associated Legendre functions. We can now write out the field
which has been quantized on the $t=0$ surface which corresponds to the geodesic
surface passing through the point $(T_0,{\bf X}_0)$ .
\begin{equation}
\Psi_1=\sum_{N=0}^{\infty}\sum_{l=0}^{N}\sum_{n=-l}^{l}\left\{
 _1a_{N l n}\phi^{(+)}_{N l n}(t,r,\theta,\phi) +  _1\! a_{N l
n}^{\dagger}\phi^{\ast(+)}_{N l n}(t,r,\theta,\phi)
\right\}
\label{3.6}
\end{equation}

\section{Particle Creation}
To investigate the particle creation in this universe, as observed by an
observer stationary with respect to the original coordinates $(T,{\bf X})$,
we calculate the Bogolubov transformation relating the annihilation and
creation operators from two different surfaces of quantization that the
observer passes through. To calculate the coefficients of this transformation
we equate the same field from two different quantizations on a common surface,
\begin{equation}
\Psi_1(t,r,\theta,\phi)=\Psi_2(t'(t,r,\theta,\phi),r'(t,r,\theta,\phi),
\theta'(t,r,\theta,\phi),\phi'(t,r,\theta,\phi)).
\label{4.1}
\end{equation}
Here $\Psi_1(t,r,\theta,\phi)$ is the field written out in
(\ref{3.6}) and
$\Psi_2(t',r',\theta',\phi')$ is the same field which has been quantized
on a second
surface $t'=0$. The ``second" field is therefore quantized for the same
observer as the first but at some later time $T'_0$ with ${\bf X}_0={\bf
X'}_0$.
All the physics of the
observations made by this observer are determined by the functions
$t'(t,r,\theta,\phi)$,
$r'(t,r,\theta,\phi)$,$\theta'(t,r,\theta,\phi)$,$\phi'(t,r,\theta,\phi)$,
and the derivatives of these functions with
respect to $t$. In this way the geometry of the spacetime via the coordinate
independent prescription we have used, determines the spectrum of created
particles. This is the reason for the comment at the end of Section II
about the form of the metric not being as important as the transformations
that gave that form of the metric. These functions take on a fairly simple
form for the stationary observer,
\begin{eqnarray}
t'&=&\frac{2}{\lambda}\sinh^{-1}\left[\sinh(\frac{\lambda
t}{2})\cosh(\tau)-\cosh(\frac{\lambda t}{2})\cos(\frac{\lambda
r}{2})\sinh(\tau) \right] \nonumber \\
r'&=&\frac{2}{\lambda}\tan^{-1}\left[\frac{\cosh(\frac{\lambda
t}{2})\sin(\frac{\lambda r}{2})}{\cosh(\frac{\lambda t}{2})\cos(\frac{\lambda
r}{2})\cosh(\tau)-\sinh(\frac{\lambda t}{2})\sinh(\tau)} \right] \nonumber \\
\theta'&=&\theta  \nonumber \\
\phi'&=&\phi \ \ \ \ \ \ {\rm where} \ \ \ \ \tau=\frac{\lambda}{2}(T'_0-T_0)
\label{4.2}
\end{eqnarray}

We calculate the Bogolubov transformation by ``matching" the
field and its first derivative with respect to $t$ at $t=0$.
This allows us to calculate the $\beta$ coefficient of the Bogolubov
transformation which gives rise to the particle creation. In calculating
the Bogolubov $\beta$ coefficient we are able to perform the $\theta$ and
$\phi$ integrals of the spherical harmonics because of the simplicity of the
coordinate transformations (\ref{4.2}) leaving,
\begin{equation}
\beta(N,N',l)=\frac{i}{2\omega_N}\int_0^\pi d\chi \sin^2(\chi) R_{Nl}(\chi)
\left(-i\omega_N f^{\ast\ (+)}_{N'}(t') R_{N'l}(\chi')+\partial_t\left(f^{\ast\
(+)}_{N'}(t')
R_{N'l}(\chi')\right)\right)|_{t=0}
\label{4.3}
\end{equation}
here $\chi=\frac{\lambda r}{2}$  and $\chi'=\frac{\lambda r'}{2}$. For
notational convenience we have split up the radial and time functions as
  \begin{eqnarray}
R_{N l}(\chi)&=& F_{N l} \sin^l(\chi){\rm C}_{N-l}^{l+1}\left(\cos(\chi)
\right)  \nonumber \\
f^{\ast\ (+)}_{N}(t') &=& {\rm sech}^{\frac{3}{2}}
(\frac{\lambda t}{2}) \left( L
P_{\frac{1}{2}+N}^{\frac{3}{2}}(\tanh(\frac{\lambda t}{2})) +  M
Q_{\frac{1}{2}+N}^{\frac{3}{2}}(\tanh(\frac{\lambda t}{2}))
\right)
\label{4.4}
\end{eqnarray}
In the next section we examine the structure of $\beta$ in detail.

\section{Total Number of Particles Created}
To show that the total number of particles created is finite we much show that
the Bogolubov transformation is Hilbert-Schmidt namely,
\begin{equation}
\sum_{N N' l}\left| \beta(N,N',l) \right|^2 < \infty.
\label{4.5}
\end{equation}
Since the sum on the left hand side of this inequality gives the number of
particles created this inequality, if it holds, implies that the total number
of particles created is finite
and that the Bogolubov transformation is unitarily implementable.
To show this  one need only be concerned with the large
$N$,$N'$ and $l$ behaviour. As the sum over $l$ is a finite
sum and $\beta(N,N',l)$ decreases with $l$ when $l$ is large
then one only need be concerned with the large $N$ and $N'$
behaviour of $\beta(N,N',l)$.
By looking at this asymptotic behaviour one is left with simpler
functions that may be integrated exactly.  We now show that
indeed when looking at the large $N$ and $N'$ behaviour the integrals
defining $\beta$ may be bounded by terms
 implying that $\left|\beta(N,N',l)\right|^2$ drops off
faster than any inverse power of $N$ and $N'$. This also implies
that the finite sum over $l$ does not change this result as it
only introduces a simple power of $N$.
 Using the following  relations
for the functions
that the modes are constructed from \cite{abram} we are able to obtain an
approximate form of $\beta(N,N',l)$ valid for large $N$ and $N'$,
\begin{eqnarray}
C^m_n[x] &=& \frac{\Gamma(2m+n)\Gamma(m+\frac{1}{2})}{\Gamma(2m)\Gamma(n+1)}
\left\{\frac{1}{4}(x^2-1)\right\}^{\frac{1}{4}-\frac{m}{2}}
 P^{\frac{1}{2}-m}_{m+n-\frac{1}{2}}(x)  \nonumber \\
P^\mu_\nu[\cos(x)] &\approx& \frac{\Gamma(\nu+\mu+1)}{\Gamma(\nu+\frac{3}{2})}
\left( \frac{1}{2}\pi \sin(x)
\right)^{-\frac{1}{2}}\cos\left((\nu+\frac{1}{2})x-\frac{\pi}{4}+\frac{\mu
\pi}{2}\right)
\ \ \ \ {\rm for\ large\ \nu}\nonumber \\
Q^m_n[\cos(x)] &\approx& \frac{\Gamma(\nu+\mu+1)}{\Gamma(\nu+\frac{3}{2})}
\left( \frac{\pi}{ 2 \sin(x)} \right)^{\frac{1}{2}}
\cos\left((\nu+\frac{1}{2})x+\frac{\pi}{4}+\frac{\mu \pi}{2}\right)
 \ \ \ \ {\rm for\ large\ \nu}\nonumber \\
\Gamma(ax+b) &\approx& \sqrt{2\pi}e^{-ax}(ax)^{ax+b-\frac{1}{2}}
\ \ \ {\rm for\ large\ a\ \ and\ x>0}.
\label{4.6}
\end{eqnarray}
The expression for $\beta$ now involves many terms but is still simple
enough to see what is required.
\begin{eqnarray}
\beta(N,N',l)&=& \int_0^{2\pi}\!\! d\chi\,\, K\left(\left( A\left(  A_1\times L
+
 A_2\times M \right)  + B\left( B_1\times L + B_2\times M \right)  \right)  M_1
\right. \nonumber \\
&+& \left.  \left( A\left(  C_1\times L +  C_2\times M \right)  +
     B\left(  D_1\times L +  D_2\times M \right)  \right)  N_1 \right)
\label{4.7}
\end{eqnarray}

where
\begin{eqnarray}
A&=& \cos ({{l\,\pi }\over 2} - \cos^{-1} ({{\cos (\chi)}\over
      {{\sqrt{{{\cos (\chi)}^2} +
           {{{\rm sech}(\tau)}^2}\,{{\sin (\chi)}^2}}}}}) -
   {\it N'}\,\cos^{-1} ({{\cos (\chi)}\over
       {{\sqrt{{{\cos (\chi)}^2} +
            {{{\rm sech}(\tau)}^2}\,{{\sin (\chi)}^2}}}}}))
  \nonumber \\
B&=& \sin ({{l\,\pi }\over 2} - \cos^{-1} ({{\cos (\chi)}\over
      {{\sqrt{{{\cos (\chi)}^2} +
           {{{\rm sech}(\tau)}^2}\,{{\sin (\chi)}^2}}}}}) -
   {\it N'}\,\cos^{-1} ({{\cos (\chi)}\over
       {{\sqrt{{{\cos (\chi)}^2} +
            {{{\rm sech}(\tau)}^2}\,{{\sin (\chi)}^2}}}}}))
	\nonumber \\
M_1&=& \cos ({\it N'} \pi  - {\it N'}
       \cos^{-1} ({{\cos (\chi){\rm sinh}(\tau)}\over
          {{\sqrt{1 + {{\cos (\chi)}^2}
                {{{\rm sinh}(\tau)}^2}}}}})  )
   \sin ({{l\pi }\over 2} - \chi -
     {\rm N}\chi )
 \nonumber \\
N_1&=& \sin ({{l\pi }\over 2} - \chi -
     {\rm N} \chi )
   \sin ({\it N'} \pi  - {\it N'}
       \cos^{-1} ({{\cos (\chi){\rm sinh}(\tau)}\over
          {{\sqrt{1 + {{\cos (\chi)}^2}
                {{{\rm sinh}(\tau)}^2}}}}})  )
 \nonumber \\
A_1&=& 16 \left( 1 + l \right)  \lambda {\it N'}
  \Gamma(2 \left( 1 + l \right) ) \Gamma({5\over 2} + l)
  \sin (\chi) {\rm sinh}(\tau)
\nonumber \\
A_2&=& -  8 \left( 1 + l \right)  \lambda {\it N'} \pi
  \cos (\chi) \Gamma(2 \left( 1 + l \right) )
  \Gamma({5\over 2} + l) \sin (\chi)
  {{{\rm sinh}(\tau)}^2}
  \nonumber \\
B_1&=&-  4 {{\cosh (\tau)}^2} \Gamma({3\over 2} + l)
  \Gamma(2 \left( 2 + l \right) )
  \left( -2 i \omega_N {{{\rm sech}(\tau)}^2} +
    2 \lambda \cos (\chi) \tanh (\tau)
   \right. \nonumber \\
&+& \left. l \lambda \cos (\chi) \tanh (\tau) +
    \lambda {\it N'} \cos (\chi) \tanh (\tau) -
    2 i {{\cos (\chi)}^2} \omega_N
     {{\tanh (\tau)}^2} \right)   \nonumber \\
B_2&=& 2 \pi  \Gamma({3\over 2} + l) \Gamma(2 \left( 2 + l \right) )
  \left( -2 \lambda \cosh (\tau) - \lambda {\it N'} \cosh (\tau) -
    2 i \cos (\chi) \omega_N {\rm sinh}(\tau)
   \right. \nonumber \\
&+& \left. l \lambda {{\cos (\chi)}^2} \cosh (\tau)
     {{{\rm sinh}(\tau)}^2} - 2 i {{\cos (\chi)}^3}
     \omega_N {{{\rm sinh}(\tau)}^3} \right)  \nonumber \\
C_1&=&- 16 \left( 1 + l \right)  \lambda {\it N'}
  \cos (\chi) \Gamma(2 \left( 1 + l \right) )
  \Gamma({5\over 2} + l) \sin (\chi)
  {{{\rm sinh}(\tau)}^2}
 \nonumber \\
C_2&=& -  8 \left( 1 + l \right)  \lambda {\it N'} \pi
  \Gamma(2 \left( 1 + l \right) ) \Gamma({5\over 2} + l)
  \sin (\chi) {\rm sinh}(\tau)
  \nonumber \\
D_1&=& 4 \Gamma({3\over 2} + l) \Gamma(2 \left( 2 + l \right) )
  \left( -2 \lambda \cosh (\tau) - \lambda {\it N'} \cosh (\tau) -
    2 i \cos (\chi) \omega_N {\rm sinh}(\tau)
   \right. \nonumber \\
&+& \left. l \lambda {{\cos (\chi)}^2} \cosh (\tau)
     {{{\rm sinh}(\tau)}^2} - 2 i {{\cos (\chi)}^3}
     \omega_N {{{\rm sinh}(\tau)}^3} \right)  \nonumber \\
D_2&=&  \pi \Gamma({3\over 2} + l)\Gamma(2\left( 2 + l \right) )
  \left( -3 i \omega_N +
    i \cos (2 \chi)\omega_N -
    i \cosh (2\tau)\omega_N -
    i \cos (2 \chi)\cosh (2 \tau) \omega_N
\right. \nonumber \\
&+& \left.2\lambda\cos (\chi){\rm sinh}(2\tau) +
    l\lambda\cos (\chi){\rm sinh}(2\tau) +
    \lambda{\it N'}\cos (\chi){\rm sinh}(2\tau)
     \right)
  \nonumber \\
K&=&{{{2^{2 l}} {\sqrt{1 + {\rm N}}} {\sqrt{{\it N'}}} {\sqrt{1 + {\it N'}}}
     {\sqrt{{2\over {{{\pi }^5}}}}} {{\Gamma(1 + l)}^2}
     \Gamma({3\over 2} + l)}\over
   {{{{1\over 4}}^l} {\sqrt{{\rm N}}}
     {{\Gamma(2 \left( 1 + l \right) )}^2}
     \Gamma(2 \left( 2 + l \right) )
     {{\left( 1 + {{\cos (\chi)}^2}
           {{{\rm sinh}(\tau)}^2} \right) }^{{3\over 2}}}}}
\label{4.8}
\end{eqnarray}
The exact form of the above expressions are not important to understanding the
large $N$ and $N'$ behaviour of $\left|\beta(N,N',l)\right|^2$. What is
important
is to notice that the expressions $A_1,A_2,B_1,B_2,C_1,C_2,D_1,D_2,K$ do not
change as far as their $N$ and $N'$ behaviour is concerned when differentiated
with respect to
$\chi$. This implies that one can integrate the expression by parts
indefinitely to observe that the expression must drop off faster than any
inverse power of $N$ and $N'$. A typical term after writing out the
trigonometric functions in terms of exponentials reads,
\begin{equation}
\int_0^{2\pi}d\chi e^{\pm iN\chi}e^{\pm iN'\cos^{-1}(p(\chi))}e^{\pm
iN'\cos^{-1}(q(\chi))}F(N,N',\chi).
\label{4.9}
\end{equation}
Here
\begin{eqnarray}
p(\chi)&=& {{\cos (\chi)}\over
       {{\sqrt{{{\cos (\chi)}^2} +
            {{{\rm sech}(\tau)}^2}\,{{\sin (\chi)}^2}}}}} \nonumber \\
q(\chi)&=& {{\cos (\chi){\rm sinh}(\tau)}\over
          {{\sqrt{1 + {{\cos (\chi)}^2}
                {{{\rm sinh}(\tau)}^2}}}}}.
\label{4.10}
\end{eqnarray}
In the above expression the exponentials represent the contributions from
the combinations of $A,B,M_1,N_1$ and $F(N,N',\chi)$ represents the
contribution
from the functions $A_1,A_2,B_1,B_2,C_1,C_2,D_1,D_2,K$.
Equation (\ref{4.9}) can be rewritten,
\begin{equation}
\int_0^{2\pi} \frac{d(e^{\pm i N \chi }e^{\pm i N'(\cos^{-1}(p)\pm \cos^{-1}(q)
)})}{\pm i N \mp i N'(\frac{1}{\sqrt{1-p^2}}\frac{dp}{d\chi}\pm
\frac{1}{\sqrt{1-q^2}}\frac{dq}{d\chi})} F(N,N',\chi)
\label{4.95}
\end{equation}
Thus an integration by parts produces a terms which drops off like,
\begin{equation}
\frac{d}{d\chi}\left(\frac{F(N,N',\chi)}{\pm i N \mp i N'(\frac{1}
{\sqrt{1-p^2}}\frac{dp}{d\chi}\pm
\frac{1}{\sqrt{1-q^2}}\frac{dq}{d\chi})}\right)
\label{4.96}
\end{equation}
Because the behaviour of $F'(N,N',\chi)$ for large $N$ and $N'$ is no worse
than $F(N,N',\chi)$ this procedure can be repeated indefinitely showing that
 $\beta(N,N',l)$ drops off faster than any inverse power of $N$ and $N'$ for
large $N,N'$. We can
then conclude that the particle creation is finite and that the Bogolubov
transformation is unitarily implementable.

Concerning the $l$ dependence in $\beta(N,N',l)$
we only have a finite sum for the total particle creation.
It is easy to show that if one uses the
same approximations (\ref{4.6}) for the gamma functions involving the $l$'s
which are valid for large $l$, $\beta(N,N',l)$ drops off for large $l$ as $l$
increases.
Thus, the probability of finding particles created with angular momentum $l$
decreases as $l$ increases. This  means that when  one does the finite sum over
$l$  the result will not grow any quicker than $N$. Therefore
because the particle density in $N$ and $N'$ drops off faster than any inverse
power of $N$ and $N'$ the total particle creation remains finite.

\section{Conclusions}
 We have calculated explicitly the particle creation observed by an observer
which is stationary in $3+1$ de Sitter space.
We calculate this particle creation by calculating the Bogolubov transformation
relating the annihilation and creation
operators from two different quantizations. These different quantizations are
constructed using the same procedure on
two different spacelike surfaces. Physically this particle creation can be
understood as the particle creation seen
by an observer moving from one of these surfaces to the next. By looking at the
large momenta behaviour for the
Bogolubov transformations we are able to show that the transformation is
unitarily implementable and therefore the
particle creation is finite. Because $\beta(N,N',l)$ drops off faster than any
inverse power of $N$ and $N'$ it may be that it drops as an exponential
suggesting a thermal spectrum.

It should be emphasized what this calculation is not a calculation of. Many
calculations have been done
calculating the Bogolubov transformations relating the creation and
annihilation operators due to two different
coordinatizations of similar spacetimes. One coordinatization usually covering
the entire spacetime and the other only
covering a portion of the spacetime. These calculations seem to require the
observers to have a split personality,
so that at one time they think they are in a geodesically complete spacetime
but at the
same time are in only a portion of
the spacetime. The procedure advocated in this paper requires that one use the
geodesically complete
coordinatization as the spacetime is spanned by geodesics in the preferred
coordinates. In this particular example this means that that the prefered
coordinatization is compact. It is this compactness that allows us to intergate
by parts the expression for the total particle creation and show it is finite.

In spacetimes where there
is a boundary present such as an horizon one may have to impose boundary
conditions at the horizon \cite{earlier}. In fact comparing coordinatizations
where one coordinatization implies a boundary and therefore does not cover the
entire manifold has been investigated in a clear paper by  Salaev and Krustalev
\cite{russ}. In this paper the authors conclude that either one has a boundary
in the spacetime or one does not, there is no in between. This is the reason
for the split personality analogy made above.

 The alternative to the split personality scenario is that the observer somehow
moves from one spacetime to
the other, an issue that has been addressed earlier by Massacand and
Schmid \cite{eth}  and argued to be unreasonable.

\section{Acknowledgements}
This research was supported in part by a grant from the Natural Sciences and
Engineering Research Council of Canada.

\end{document}